# PTAL multi-spectral database of planetary terrestrial analogues: Raman data overview

*Marco Veneranda[1], Jesus Saiz[1], Aurelio Sanz-Arranz[1], Jose Antonio Manrique[1], Guillermo Lopez-Reyes[1], Jesus Medina[1], Henning Dypvik[2], Stephanie C. Werner[2], Fernando Rull[1]*

[1] *Department of Condensed Matter Physics, Crystallography and Mineralogy, Univ. of Valladolid, Spain. Ave. Francisco Vallés, 8, Boecillo, 47151 Spain.* marco.veneranda.87@gmail.com
[2]*Department of Geosciences,* CEED/GEO*, University of Oslo, Norway*

## Abstract

The multi analytical study of terrestrial analogues is a useful strategy to deepen the knowledge about the geological and environmental evolution of Mars and other extraterrestrial bodies.
In spite of the increasing importance that LIBS, NIR and Raman techniques are acquiring in the field of space exploration, there is a lack web-based platform providing free access to a wide multi-spectral database of terrestrial analogue materials.
The Planetary Terrestrial Analogue Library (PTAL) project aims at responding to this critical need by developing and providing free web accessibility to LIBS, NIR and Raman data from more than 94 terrestrial analogues selected according to their congruence with Martian geological contexts. In this framework, the present manuscript provides the scientific community with a complete overview of the over 4500 Raman spectra collected to feed the PTAL database.
Raman data, obtained through the complementary use of laboratory and spacecraft-simulator systems, confirmed the effectiveness of this spectroscopic technique for the detection of major and minor mineralogical phases of the samples, the latter being of critical importance for the recognition of geological processes that could have occurred on Mars and other planets.
In light of the forthcoming missions to Mars, the results obtained through the RLS ExoMars Simulator offer a valuable insight on the scientific outcome that could derive from the RLS spectrometer that will soon land on Mars as part of the ExoMars rover payload.

**Keywords:** PTAL project; Terrestrial Analogues; Raman Spectroscopy; RLS ExoMars Simulator; Mars;

### 1 Introduction

Spectroscopic techniques are acquiring an increasing importance in the field of Mars exploration. For example, being part of the ChemCam instrument package of the Curiosity/NASA rover, a remote Laser Induced Breakdown Spectroscopy (LIBS) system is shedding light since 2012 on the geochemical composition of Martian soils and rocks.[1] In light of the outstanding results provided by ChemCam, a more advanced LIBS system is scheduled to deploy on Mars in 2020 through the Mars2020/NASA mission[2], featured by the SuperCam instrument.

With regards to molecular spectroscopic techniques, since 2003 the Mars Express/ESA orbiter is mapping the mineral composition of the Martian surface through the OMEGA (Observatoire pour la Minéralogie, l'Eau, les Glaces et l'Activité) visible-infrared spectrometer. [3] From 2006, OMEGA analysis are flanked by CRISM data (Compact Reconnaissance Imaging Spectrometer for Mars), the visible-infrared system aboard the Mars Reconnaissance Orbiter.[4] The hyperspectral information obtained from OMEGA and CRISM will be soon complemented by detailed data in a microscopic scale thanks to the MicrOmega VIS/NIR spectrometer[5] that will deploy on Mars in 2020 as part of the ExoMars-Pasteur payload.

Even though Raman spectroscopy has an uncountable number of terrestrial applications, it has never been employed in space so far. However, the ExoMars/ESA rover will soon deploy on Mars the Raman Laser Spectrometer (RLS) system, the first Raman instrument in history to be validated for space exploration missions.[6] This mission, with its highest aim being the detection of organic materials or biosignatures on mineralogical samples collected down to a depth of two meters,[7] features the Mars Organic Molecule Analyzer (MOMA) in addition to MicrOmega and RLS as instruments of the Analytic Laboratory Drawer (ALD) of the rover. Besides ExoMars, the payload of the Mars2020's rover will include two different Raman systems as parts of Sherloc[8] and SuperCam[9] instrument packages.

Beyond Mars exploration, the use of Raman spectroscopy has been proposed for missions to further extraterrestrial bodies. This is the case of the Europa Lander/NASA proposal, which is planning to explore the sub-surface of Europe (Jupiter's satellite) in search of potential biosignatures.[10]

Considering the increasing weight these techniques are acquiring in the field of planetary space exploration, the spectroscopic characterization of terrestrial analogues become an essential tool to estimate the potential scientific return of the forthcoming ESA and NASA missions.[11,12]

In the past years, several works have been published on this issue.[13-17] For example, M. Veneranda et al.[18] made use of NIR, LIBS and Raman techniques to characterize a breccia sample selected from the Chesapeake Bay Impact Structure (CBIS). The results obtained from the use of the RLS ExoMars Simulator, which effectively detected shocked quartz and minor hydrothermal-related phases demonstrate the potential key role of the ExoMars/RLS system in the discrimination of wet-target craters on Mars.

In spite of the crucial role played by the spectroscopic study of terrestrial analogues, three main problems must be underlined: 1) the results obtained so far in this field of research are still dispersed across the scientific literature, 2) there is a lack of accessible database of terrestrial analogues analysed through flight-derived analytical instruments, and 3) many studies are carried out using spectroscopic systems providing analytical performances not comparable to those reachable by spacecraft instrumentation.

To overcome this limitation, the overall purpose of the Planetary Terrestrial Analogues Library (PTAL) project[19] is to enhance the scientific outcomes that could derive from the study of terrestrial analogues by providing to scientific community an extensive multi-spectral database of terrestrial analogues. For this purpose, both laboratory and spacecraft-simulator instrumentations were used to analyse: 1) a large collection of natural geological samples collected from terrestrial analogues sites and 2) artificial samples replicating Martian protoliths composition and altered in the laboratory under controlled physical-chemical conditions.

Raman data, together with the complementary NIR and LIBS analysis, will be supported by XRD data and thin section observations so as to provide an exhaustive geochemical and mineralogical characterization of the selected samples.

In the framework of the PTAL project, the present article aims at presenting an overview of the whole set of Raman spectra collected from the 94 PTAL samples through the complementary use of a laboratory spectrometer and the RLS ExoMars Simulator. In light of the future ExoMars mission, the comparison between the results obtained from the two systems provides valuable insight about the scientific outcome that could derive from the RLS that will soon operate on Mars.

## 2 Material and methods

### *2.1 Terrestrial analogues*

The 94 natural analogues samples selected for the PTAL database can be divided in two main categories. On one hand, unaltered geological samples were chosen according to their congruence to Martian precursor materials. Beside their characterization, those samples will be used in the future for laboratory alteration experiments. On the other hand, altered materials were selected to be used as analogues of the weathered rocks and soils that are expected to be found on Mars.[19]

In this work the geological context of the PTAL terrestrial analogue sites has been just introduced, given that the detailed description of the selected samples will be provided in more specific publications. The collected samples can be grouped in three sub-sections according to their chemical and mineralogical features: 1) hydrate sulphate minerals, 2) Martian protoliths analogues, and 3) impact craters.

Taking into account that the PTAL website will provide the opportunity to request physical access to samples, up to 1.3 kg of material was sampled for each analogue. In this way, sample accessibility will enable future users to combine PTAL spectroscopic data with further laboratory analysis. In order to meet the requirements of each analytical technique included in the PTAL project, small fragments were selected from each sample and milled at different granulometries. Coarse powder with a grain size up to 500 µm (optimal for Raman and NIR analysis, also somehow resembling the ExoMars rover crushing grain size range) was prepared using a sling mill. Besides, fine-powder with a grain size below 150 µm (optimal for XRD analysis) was obtained by further crushing the coarse powder in an agate mill (McCrone Micronizer Mill) during 12 minutes.

### *2.2 Raman systems*

The overall molecular characterization of powdered samples was performed through a Raman spectrometer assembled in the laboratory. The instrument is composed of the following commercial components: a Research Electro-Optics LSRP-3501 excitation laser (Helium-Neon) emitting at 633 nm, a KOSI Holospec1.8i spectrometer and an Andor DV420A-OE-130 CCD detector. The Nikon Eclipse E600 microscope coupled to the system is equipped with interchangeable long WD objectives of 5x, 10x, 20x, 50x and 100x. In this case, between 15 and 30 spectra were collected from each sample by focusing the excitation laser on the most interesting crystals. In this work, the use of 50x and 100x objectives was interchanged depending on the size of the analysed mineral grain. The Hologram 4.0 software was used to collect Raman spectra in the range between 130 and 3780 $cm^{-1}$.

Further Raman analysis were then carried out by means of the so-called RLS ExoMars Simulator. As reported elsewhere [18] this system, developed by the UVa-CSIC-CAB Associated Unit ERICA (Spain), is considered one of the most reliable tools to effectively emulate the scientific outcome that will be potentially produced by RLS instrument on Mars. Indeed, the hardware components of the instrument ensure analytical performances qualitatively congruent with those of the ExoMars/RLS system. The instrument is composed of a BWN-532 excitation laser (B&WTek) emitting at 532 nm, a BTC162 high resolution TE Cooled CCD Array spectrometer (B&WTek) and an optical head (B&WTek) with a long WD objective of 50x. Furthermore, the vertical and horizontal positioners coupled to the RLS ExoMars Simulator have been designed to emulate the original Sample Preparation and Distribution System (SPDS) of the ExoMars rover. Finally, the operations and algorithms carried out by the instrument to autonomously obtain high quality spectra (i.e. integration time/number of accumulations selection and signal to noise ratio optimization) are the ones designed for the RLS on Mars.[20] In this case, coarse powdered samples were analysed by simulating the operational conditions required by the Analytical Laboratory Drawer (ALD) of the ExoMars rover. Concretely between 20 and 40 spots of analysis were automatically selected from each powdered sample by moving the positioners in the x-axis at steps of 100 microns. To collect optimal spectra, the RLS ExoMars Simulator automatically performed autofocus and the the signal to noise optimization by adjusting the integration time and number of accumulations at each spot. The LabVIEW 2013 software (National Instruments, EEUU) was used to collect Raman spectra in the range between 70 and 4200 $cm^{-1}$.

### _2.3 Data treatment and interpretation_

Treatment and interpretation of the whole set of Raman data was carried out through the IDAT/SpectPro tool. As reported elsewhere,[21,22] the Instrument Data Analysis Tool (IDAT) software was developed in the framework of the ExoMars mission for the reception, decodification and verification of the scientific and housekeeping data that will be generated by the RLS instrument on Mars. As part of the IDAT software, the SpectPro application will provide the ExoMars scientific team with the whole set of tools needed for the visualization, treatment and interpretation of Raman spectra collected by the RLS on Mars. Though developed in the framework of the RLS instrument development, this software tool will potentially be usable for spectroscopic data of all types.

IDAT/SpectPro software has been programmed by the UVa-CSIC-CAB Associated Unit ERICA, which is also leading, together with INTA, the development of the RLS/ExoMars Raman spectrometer. Being part of the PTAL consortium, the ERICA research group will facilitate to future PTAL users a direct access from SpectPro to the database, using the same credentials for access to the PTAL web interface. The free access to the algorithms developed for data treatment and visualization will provide the scientific community with the set of functions necessary for a refined interpretation of the spectroscopic data included in the database (i.e. SNR calculation, baseline removal, filtering, cutting, band adjustment and many other spectral operations and the definition and use of automated routines).

The treatment of all the spectra summarized in this work was performed by using the SpectPro tool so as to provide the scientific community and future PTAL users with an insight of its potential.

**3 Results**

The interpretation of the over 4500 Raman spectra summarized in this work was mainly focused on the identification of the mineral groups composing the analysed analogue materials (i.e. feldspar, pyroxenes, olivines). Indeed, the PTAL project foresees to promote the use of SpectPro software by the users that will be able to carry out a more refined spectra interpretation by using the SpectPro data processing tools and by performing comparative analysis with standard spectra.

As detailed below, the PTAL samples are organized in three main categories:

3.1 Hydrate sulphate minerals

The mineralogical data collected by the XRD system on board of the Opportunity/NASA rover proved the presence of jarosite ($KFe_3(SO_4)_2(OH)_6$) on Meridiani Planum.[23] The identification of hydrous sulphate minerals is one of the strongest evidences in support of the past presence of water on Mars. In this light, the PTAL database includes terrestrial analogue samples collected from geological sites rich in hydrated sulphates that will help deepen the knowledge about the conditions favouring their mineralization.

*Jaroso Ravine*

Jaroso Ravine (Spain) is part of the Jaroso Hydrothermal System (JHS), a volcanic-related hydrothermal episode that, due to the contribution of sulfuric acids, generated a great variety of oxides and sulphate minerals.[24,25] This site is where jarosite was first discovered on Earth (1852) and its study could help to better comprehend the processes behind the mineralization of this compound on Mars.

For this reason, 3 different analogue materials were selected so as to be included in the PTAL database. According to spectroscopic results, sample JA08-501 is almost completely composed of jarosite. This sulphate of potassium and iron, was clearly detected thanks to its characteristic vibrational peaks at 166, 220, 298, 353, 434, 452, 512, 570, 625, 1005, 1101 and 1154 $cm^{-1}$.[26] In addition to jarosite, a Ti-based oxide (anatase [27], $TiO_2$, main peaks at 144, 400 and 641 $cm^{-1}$) was also identified by means of RLS ExoMars Simulator analysis.

Sample JA08-502 is mainly composed of quartz (strong Raman peak at 464 $cm^{-1}$). In addition to $SiO_2$, minor amounts of goethite (FeO(OH), main peaks at 241, 299 and 387 $cm^{-1}$ [28]), carbon (broad bands around 1350 and 1585 $cm^{-1}$) and rutile (a $TiO_2$ polymorph, main peaks at 246, 443 and 610 $cm^{-1}$ [27]) were detected by means of the microRaman system, while rutile was identified through RLS ExoMars Simulator analysis.

Sample JA08-503 was collected from a shale-based soil and displays a very complex mineralogy. As shown in Table 1, quartz, hematite ($Fe_2O_3$, main peaks at 222, 295, 411, 660 and 1315 $cm^{-1}$ [28]), carbon and barite ($BaSO_4$, main peaks at 460, 617,989 and 1143 $cm^{-1}$ [29]) were detected by both Raman systems, while the additional presence of goethite, biotite ($K(Mg,Fe)_3AlSi_3O_{10}(OH,F)_2$, main peaks at 261, 423 and 700 $cm^{-1}$ [29]) and anatase were confirmed by microRaman results. In addition to those, a strong vibrational peak at 1086 $cm^{-1}$ was detected on several RLS ExoMars Simulator spectra, confirming the detection of calcium carbonate ($CaCO_3$) [29].

*Rio Tinto*

Rio Tinto (Spain) is one of the most characteristic locations within the Iberian Pyrite Belt, a geographical area were more than 20 different hydrated sulphate minerals can be found, some of which were detected on Mars.[12, 30] In spite of the strong acidity of the waters flowing in Rio Tinto (pH 1.5-3.0), clear biosignatures were detected on exposed mineral substrates.[12] From the perspective of searching for biosignatures on Mars, Rio Tinto samples could help evaluating the geological and environmental conditions leading to the proliferation of organisms in extremophilic conditions.

Three samples were collected from this analogue site so as to be analysed and included in the PTAL database. Quartz, hematite and goethite were clearly detected by both Raman system as main mineralogical phases of the sample, while the detection of weak barite signals was ensured by RLS ExoMars Simulator analysis. Regarding sample RT03-502, microRaman analysis confirmed the detection of Fe-based compounds, such as hematite and pyrite ($FeS_2$, main Raman peaks at 343, 379 and 429 $cm^{-1}$ [29]). Sample RT03-503 was collected from an area rich in sulphide-based minerals. As shown in Fig. 1, Raman results highlighted the presence of pyrite as the main mineral phase of the analogue. In addition to this iron sulphide, the minor presence of quartz was also detected by both spectrometers.

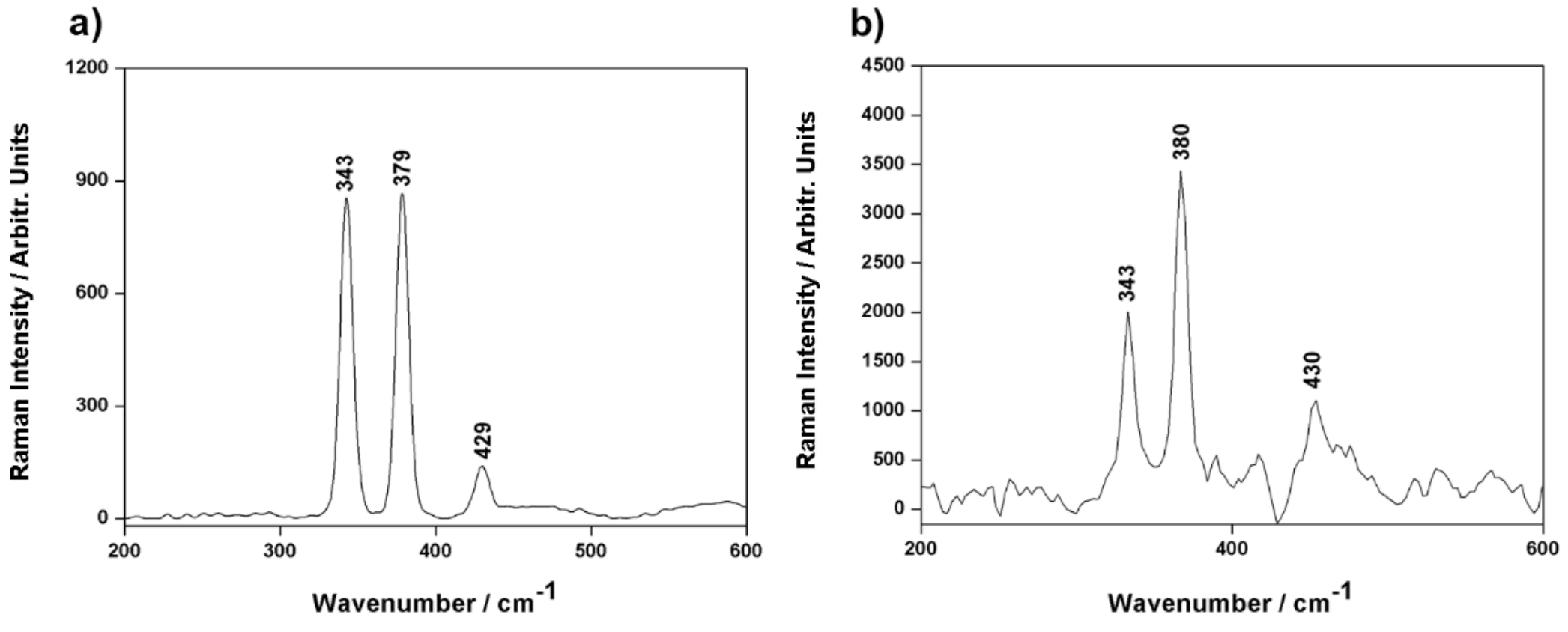

Fig. *1: Pyrite spectra collected from sample RT03-503 by means of both microRaman system (a) and RLS ExoMars Simulator (b).*

*Table 1: Overview of Raman results from Jaroso Ravine and Rio Tinto samples.*

| Sample details | | | | | | | | | | | | |
|---|---|---|---|---|---|---|---|---|---|---|---|---|
| ID code | Sampling site | Quartz | Hematite | Goethite | Jarosite | Pyrite | Biotite | Anatase | Rutile | Calcite | Carbon | Barite |
| JA08-501 | Jaroso Ravine | | | | ▩ | | | ▥ | | | | |
| JA08-502 | Jaroso Ravine | ▩ | | ▤ | | | | ▥ | ▤ | | ▤ | |
| JA08-503 | Jaroso Ravine | ▩ | ▩ | ▤ | | | ▤ | ▤ | | ▥ | ▩ | ▩ |
| RT03-501 | Rio Tinto | ▩ | ▩ | ▩ | | | | | | | | ▥ |
| RT03-502 | Rio Tinto | | ▤ | | | ▤ | | | | | | |
| RT03-503 | Rio Tinto | ▩ | | | | ▩ | | | | | | |

▩ *Both microRaman and RLS simulator* ▥ *RLS simulator* ▤ *microRaman*

## 3.2 Martian protoliths analogues

The PTAL database also includes an extensive collection of gabbros and basalts collected from analogue sites with a lithology comparable to Martian protoliths. The comprehensive characterization of these samples will play a crucial role in the selection of the optimal materials to be employed for laboratory alteration experiments. These tests, based on the exposition of the sample to controlled atmospheric conditions, will help obtaining outstanding inferences about the alteration processes that have occurred or are occurring on Mars.

### *3.2.1 Iceland*

As presented by B.L. Ehlmann et al.[31] the mineralogical and geochemical composition of Icelandic basalt rocks made them optimal analogues of basalt terrains from Noachian Mars. In the framework of the PTAL project, several rock samples were collected from different locations of the Reykjanes Peninsula (Stapafell, Krysuvik, Reykjanes and Grindavik). According to the Raman results displayed in Table 2, basalt rocks from Reykjanes, Stapafell and Grindavik are mainly composed of olivine ($(Mg,Fe)_2SiO_4$, main peaks around 822 and 852 $cm^{-1}$ [27]), pyroxene (general formula $XY(Si,Al)_2O_6$, where X can be Na, Ca, Mn, Fe, Mg or Li, while Y can be Mn, Fe, Mg, Al, Cr and Ti. Main peaks around 660 and 1010 $cm^{-1}$ [27]) and feldspars (general formula $XYNaAlSi_3O_8$, main Raman peaks at 510 and 474 $cm^{-1}$ [32]).

As can be observed on Fig. 2, the Raman spectra of olivine slightly changes depending on the crystal grain under analysis. Indeed, the main peaks around 822 and 852 $cm^{-1}$ shift towards lower or higher wavenumbers depending on the concentration ratio between fayalite ($Fe_2SiO_4$) and forsterite ($Mg_2SiO_4$) respectively.

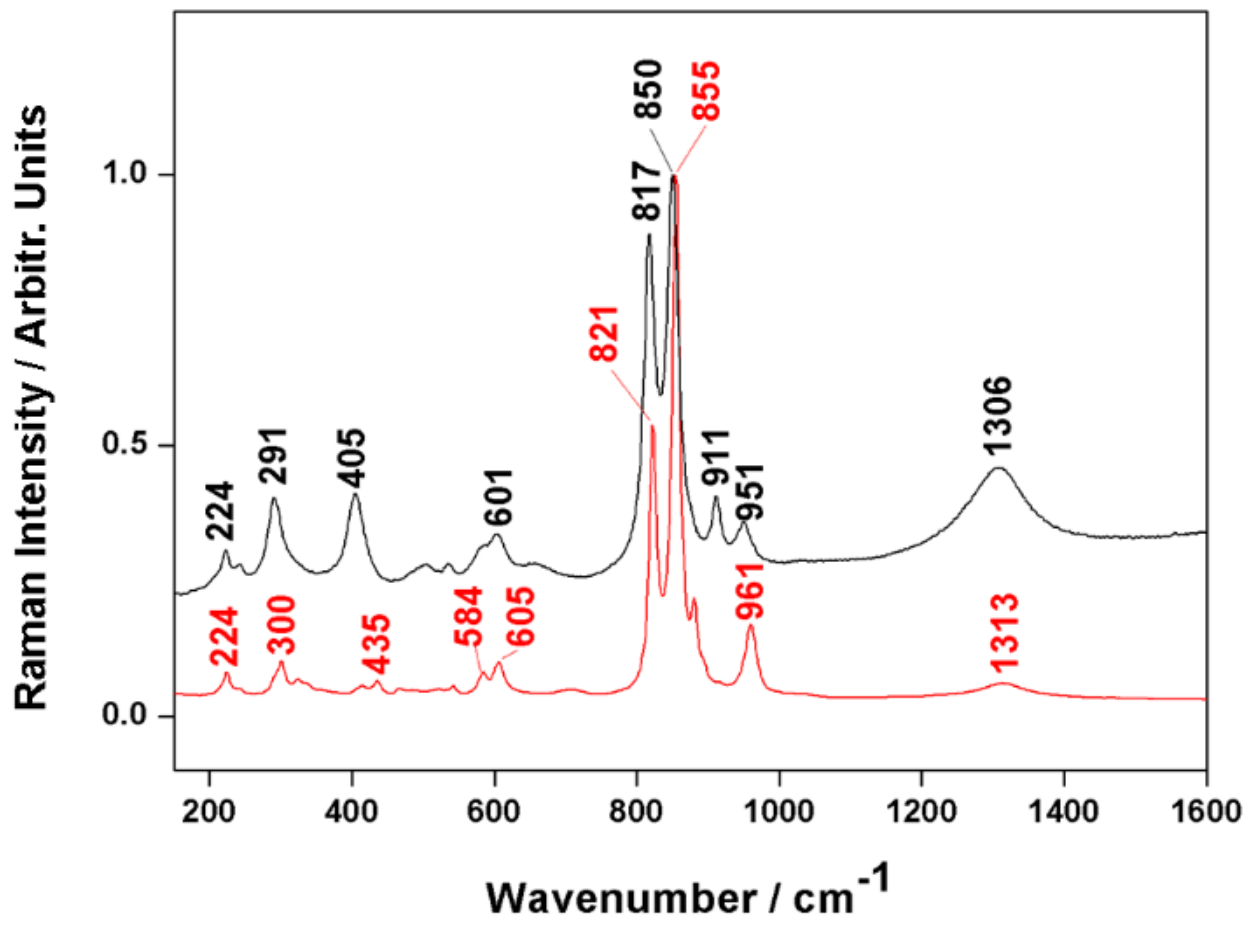

*Fig. 2: Different olivine spectra collected from sample IS16-0001 by means of the microRaman system.*

Among basalt rocks, 7 samples display alteration products in form of iron oxides (hematite on samples IS16-0001, -0002, -0004, -0005, -0011 and -0014, goethite on samples IS16-0005 and -0016). Further alteration products were also detected, as in the case of pyrite and calcite in sample IS16-0002, and serpentine in sample IS16-0007. Considering that, as explained by J. Delvigne et al.,[33] olivine have the highest value of weathering potential index among silicate minerals, the systematic detection of olivine and sporadic identification of alteration products proves that the selected materials have a weak or null degree of alteration. Regarding the presence of additional minor phases, it must be also underlined the detection of ilmenite in sample IS16-0001, -0002 and -0008, as well as the detection of glass (broad bands around 450 and 1100 $cm^{-1}$) in samples IS16-0006 and -0009.

Besides basalt rocks, the Icelandic collection also includes three solfatara samples, named IS16-0010, IS16-0011 and S16-0012. The collected materials represent the precipitation product of the mixture of sulfuric gas, carbon dioxide and steam emitted in the Krysuvik geothermal area. According to Raman results, the bright white color of samples IS16-0010 and IS16-0011 is related to the detection of anatase, while the yellow appearance of sample IS16-0012 is attributed to the high concentration of sulphur (Fig. 3, main peaks at 183, 219 and 473 $cm^{-1}$[29]).

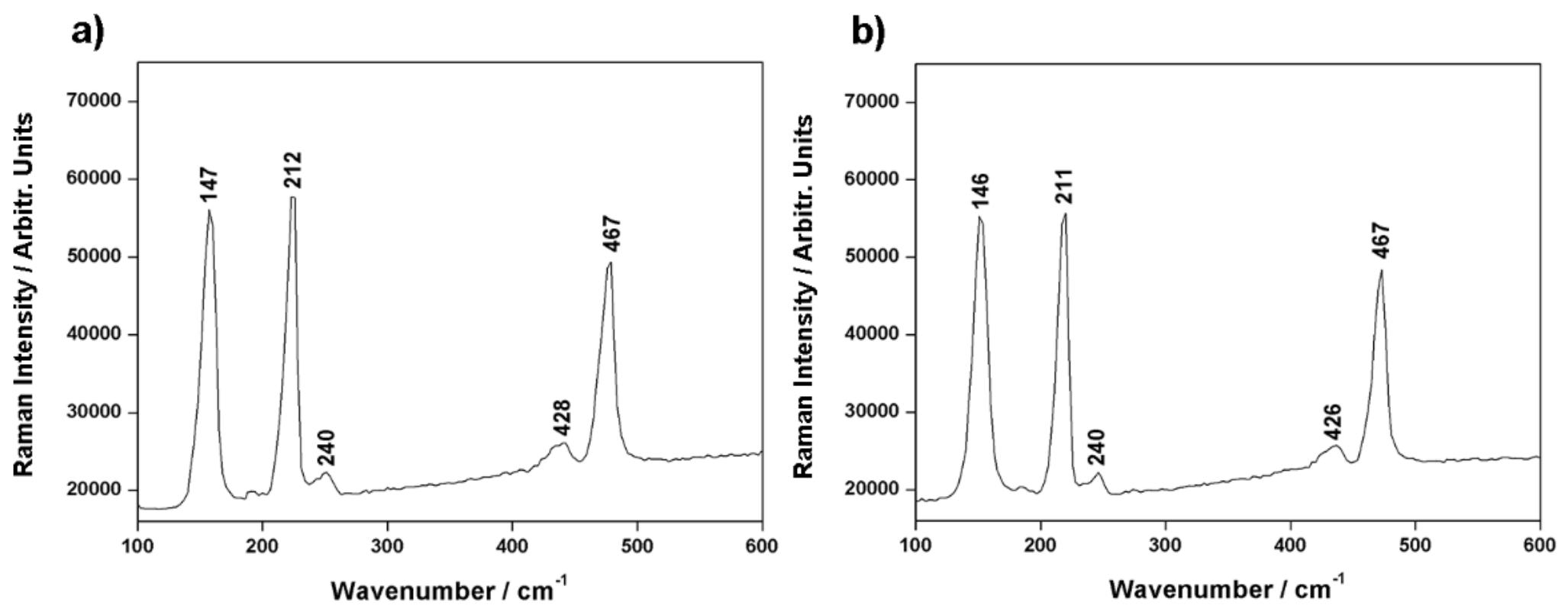

*Fig. 3: Sulphur spectrum collected from sample IS16-0012 by means of both microRaman system (a) and RLS ExoMars Simulator (b).*

*Table 2: Overview of Raman results collected from Icelandic samples.*

| **Sample details** | | **Detected compounds** | | | | | | | | | | | | | |
|---|---|---|---|---|---|---|---|---|---|---|---|---|---|---|---|
| ID code | Sampling site | Feldspar | Pyroxene | Hematite | Goethite | Olivine | Serpentine | Apatite | Anatase | Calcite | Pyrite | Sulphur | Carbon | Glass | Ilmenite |
| IS16-0001 | Reykjanes | ▩ | ▩ | ▩ | | ▩ | | | | | | | | | ▤ |
| IS16-0002 | Reykjanes | ▩ | ▩ | ▩ | | ▩ | | | | ▤ | ▤ | | | | ▤ |
| IS16-0003 | Stapafell | ▩ | ▩ | | | ▩ | | | | | | | | | |
| IS16-0004 | Stapafell | ▩ | ▩ | ▤ | | ▩ | | | | | | | | | |
| IS16-0005 | Stapafell | ▩ | ▩ | ▤ | ▥ | ▩ | | | | | | | | | |
| IS16-0006 | Stapafell | | | | | ▤ | | | | | | | | ▤ | |
| IS16-0007 | Stapafell | ▩ | ▩ | | | ▩ | ▤ | | | | | | | | |
| IS16-0008 | Stapafell | ▩ | ▩ | | | ▩ | | | | | | | | | ▤ |
| IS16-0009 | Stapafell | | ▥ | | | ▥ | | | | | | | | ▤ | |
| IS16-0011 | Krysuvik | | ▤ | ▤ | | | | | ▥ | | ▤ | ▤ | | | |
| IS16-0012 | Krysuvik | | | | | | | | ▥ | | | ▩ | | | |
| IS16-0013 | Reykjanes | ▩ | ▩ | | | ▩ | | | | | | | | | |
| IS16-0014 | Reykjanes | ▤ | ▤ | ▤ | | | | ▥ | | | | | | | |
| IS16-0015 | Grindavik | ▩ | ▩ | | | ▩ | | | | | | | ▥ | | |
| IS16-0016 | Grindavik | ▩ | ▤ | | ▤ | ▩ | | | | | | | | | |

▩ *Both microRaman and RLS simulator* ▥ *RLS simulator* ▤ *microRaman*

### *3.2.2 Canary Islands*

The Canary archipelago (Spain) is composed of 7 main islands and several islets formed from oceanic intraplate volcanism activity. Among them, Grand Canary and Tenerife islands show the so-called "azulejos" alteration of basalts, a hydrothermal process that takes place when fresh erupted magma enters in contact with seawater (hydromagmatic eruption). In the light of space exploration, to comprehend the mineralogical transformations triggered by this alteration process could help discriminating potential hydromagmatic eruptions that might have occurred on Mars and other extraterrestrial bodies.[13]

#### *Grand Canary*

The Grand Canary island is considered a unique case of study since it displays 4 different "azulejos" alteration strata separated from each other by unaltered pyroclastic deposits.[34,35]

Thus, fresh, weathered, and hydrothermal altered ("azulejos") basalts were collected from different locations of Grand Canary and analysed for comparison purposes.

As displayed in Table 3, the mineralogical composition of not hydrothermal-altered is dominated by feldspars and pyroxenes, being those, the major phases of basalt rocks. Raman analysis of the Agaete (AG16-0001) and Roque Nublo (RN16-0001) samples detected additional basalt minerals such as olivine (as major phase) and apatite, while no alteration products were identified. Besides, sandstone sample from Bc. Tamaraceite (BT16-0001) and pumice from Pica Bandama (CB16-0001) also includes titanium-based oxides, such as ilmenite ($FeTiO_3$, main peak at 685 $cm^{-1}$) and anatase. By using the RLS ExoMars Simulator, the characteristic vibrational peaks of calcium carbonate were clearly detected on sample CB16-0001, corroborating a light alteration (carbonatation process) of the sample. In addition to feldspar and pyroxene crystals, optical images collected from sample TO16-0001 (Punta Camello) by means of the microRaman microscope enabled the detection of additional grains with a very intense blue colour. As represented in Fig. 4, the vibrational profile provided by blue crystals (main peaks at 257, 289, 546, 582, 804, 989, 1093 and 1350 $cm^{-1}$) fit with lazurite,[29] a rare tectosilicate mineral with formula $Na_3Ca(Al_3Si_3O_{12})S$.

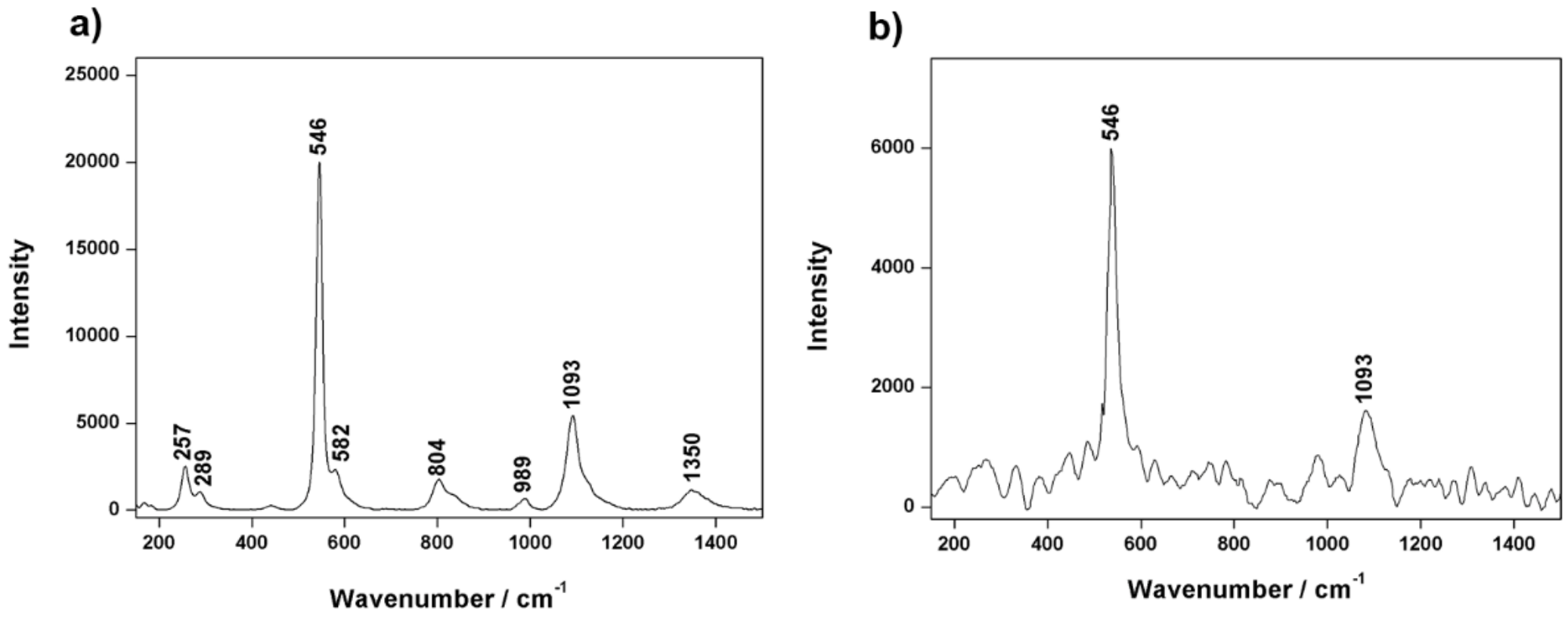

*Fig. 4: Lazurite spectra collected from sample TO16-0001 by means of both microRaman system(a) and RLS ExoMars Simulator (b).*

With regards to the "azulejos" altered rocks, the results displayed in Table 3 clearly highlight that the mineralogical composition of samples FA16-0001, FA16-0002 and FA16-0003 is very different from the materials previously described. Indeed, pyroxenes are not present while quartz, hematite and anatase were clearly detected together with feldspars (major compound). In the case of sample FA16-0001, dolomite ($CaMg(CO_3)_2$, main peak at 1096 $cm^{-1}$ [29]) and calcite were also detected, proving the occurrence of intense carbonatation processes.

*Table 3: Overview of Raman results collected from Grand Canary samples.*

| Sample details | | Detected compounds | | | | | | | | | | |
|---|---|---|---|---|---|---|---|---|---|---|---|---|
| ID code | Sampling site | Feldspar | Pyroxene | Quartz | Hematite | Olivine | Apatite | Dolomite | Calcite | Anatase | Lazurite | Ilmenite |
| AG16-001 | Agaete | ▩ | ▩ | | | ▩ | ▩ | | | | | |
| TO16-001 | Punta Camello | ▩ | ▤ | | | | | | | | ▩ | |
| BT16-001 | Bc.Tamaraceite | ▩ | ▩ | ▤ | | | ▤ | | | ▥ | | ▤ |
| BT16-002 | Bc.Tamaraceite | ▥ | ▤ | | | ▩ | | | | | | |
| CB16-001 | Pica Bandama | ▥ | ▩ | | | ▩ | ▥ | | ▥ | | | ▤ |
| FA16-001 | Fuente de los Azulejos | ▩ | | ▩ | ▤ | | | ▩ | ▩ | ▩ | | ▤ |
| FA16-002 | Fuente de los Azulejos | ▩ | | ▩ | ▩ | | | | | ▩ | | |
| FA16-003 | Fuente de los Azulejos | ▥ | | ▩ | ▤ | | | | | ▥ | | |
| RN16-001 | Roque Nublo | ▤ | ▩ | | | ▩ | ▤ | | | | | |

▩ *Both microRaman and RLS simulator*  ▥ *RLS simulator*  ▤ *microRaman*

*Tenerife*

As in the case of Grand Canary, further analogue samples were collected from the older basalt outcrops (Upper Miocene-Upper Pleistocene) of Tenerife island (both unaltered and altered in hydrothermal conditions).[14]

As shown in Table 4, all analogues not related to “azulejos” alteration are mainly composed of feldspars and pyroxenes. A detailed analysis of the Raman results obtained from the samples collected from “Mina Reventada” showed that olivine and apatite were also detected as minor phases. Furthermore, by means of the MicroRaman system, additional minerals as hematite and ilmenite were detected on sample MR16-0001 and MR16-0002 respectively.

The mineralogical composition of Adeje analogues is similar to the sample from Mina Reventada. However, in the case of sample AD16-0001, the absence of olivine and the presence of calcite and amorphous carbon is to be noted. The latter, detected thanks to its characteristic bands at 1350 and 1585 $cm^{-1}$, indicates organic material inclusions within the lithological sample.

Two samples of sandstone were collected from the Amarilla mountain, located in the southern area of Tenerife island. In this case, feldspars, pyroxenes, hematite and ilmenite were detected by both spectrometers as main mineral phases of the two analogues.

Four altered samples (TF16-0002, TF16-0028, TF16-0059 and TF16-0066) were selected from Azulejos, a locality that takes its name from the hydrothermal process itself. In this case pyroxenes were not observed, while anatase (Fig. 5) was the only detected titanium-based mineral (instead of ilmenite). Calcium carbonate was identified on both TF16-0059 and TG16-0066, while apatite was only observed on sample TF16-0059. The composition of “azulejos” samples from Tenerife and Grand Canary fit very well, clearly indicating the effects triggered by hydrothermal alteration process in the mineralogy of both locations.

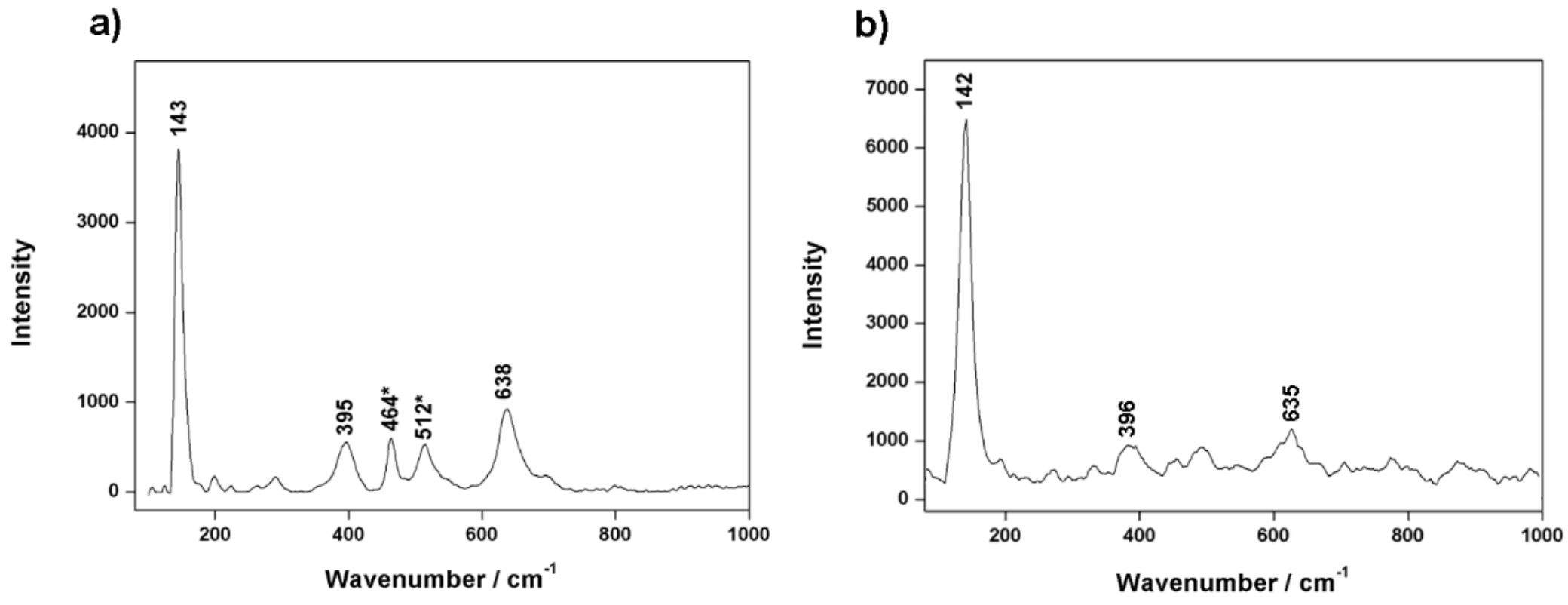

*Fig. 5: Anatase spectra collected from sample TF16-0002 by means of both microRaman system (a) and RLS ExoMars Simulator (b).*

*Table 4: Overview of Raman results from Tenerife samples.*

| Sample details | | Detected compounds | | | | | | | | | |
|---|---|---|---|---|---|---|---|---|---|---|---|
| ID code | Sampling site | Feldspar | Pyroxene | Quartz | Hematite | Olivine | Apatite | Anatase | Calcite | Carbon | Ilmenite |
| MR16-001 | Mina Reventada | ▩ | ▩ | | ▤ | ▤ | ▩ | | | | |
| MR16-002 | Mina Reventada | ▩ | ▩ | | | ▤ | ▩ | | | | ▤ |
| AD16-001 | Adeje | ▩ | ▤ | | ▤ | | ▩ | | ▥ | ▤ | ▤ |
| AM16-001 | Montaña Amarilla | ▩ | ▩ | | ▤ | | | | | | ▤ |
| AM16-002 | Montaña Amarilla | ▩ | ▥ | | ▩ | ▥ | ▥ | | | | ▩ |
| TF16-002 | Los Azulejos | ▤ | | | ▤ | | | ▩ | | | |
| TF16-028 | Los Azulejos | ▤ | | ▥ | | | | ▩ | | | |
| TF16-059 | Los Azulejos | ▤ | | ▤ | ▤ | | ▥ | ▩ | ▤ | | |
| TF16-066 | Los Azulejos | ▩ | | | ▩ | | | ▩ | ▤ | | |

▩ *Both microRaman and RLS simulator* ▥ *RLS simulator* ▤ *microRaman*

### *3.2.3 John Day Formation*

The John Day Formation displays a thick series of rock strata deposited between 37 and 20 million years ago from the massive fall of ash and dust emitted by a volcanic regime linked to tectonic activity.[36] The different colours of the geological layers rely in the deposition of volcanic materials of different elemental and molecular composition. The zeolitic diagenesis of these layers makes the John Day formation an optimal terrestrial analogue to evaluate the capability of spectroscopic techniques to detect $H_2O$ and/or biomarkers in the interlayer spaces of clay particles,[37] from which important insight about the past presence of water and/or life on Mars can be inferred.[38] Furthermore, it could be useful in the detection of minor and trace compounds, from which important inference about the mineralogy evolution of the site can be obtained.

As displayed on Table 5, clay samples selected from John Day formation include feldspars, hematite (Fig. 6) and anatase as minor or trace compounds. By comparing the analytical results obtained from the two Raman systems, it can be noted their different sensitivity towards the detection of hematite and anatase. On one hand, hematite can be easily detected by the microRaman thanks to the so-called resonance Raman effect that take place when the wavelength of the excitation laser is close to the electronic transition of the compound under analysis. On the other hand, the technical features of the RLS ExoMars Simulator make it capable of detecting the main vibrational peak of anatase, which appears in a range of wavelength that cannot be measured by microRaman (144 $cm^{-1}$).

Beside feldspars, hematite and anatase, additional minor compounds were detected. For example, analogue materials collected from Foree are distinguished by the absence of hematite and feldspars, and the additional presence of calcium carbonate.

Among the selected samples, the ones collected from Picture Gorge (JD16-0004, JD16-0005 and JD16-0007) are the only ones clearly displaying the vibrational features of pyroxenes. In the case of samples JD16-0004 and JD16-0007, the further detection of clinoptilolite ($(Ca,\underline{K,Na})_6(Si_{30}Al_6)O_{72}\cdot 20H_2O$, main peaks at 405 and 483 $cm^{-1}$ [29]) and olivine was confirmed respectively. The mineralogical composition of Picture Gorge samples fits with the visual mineralogical interpretation provided by geologists, who classified the collected materials as altered and unaltered basalts.

Several Raman spectra from Painted hills analogues also showed the vibrational bands of carbon around 1350 and 1585 $cm^{-1}$. Similarly, the samples selected from Hancock field station are the only ones (together with JD16-0002 and JD16-0015) showing quartz as minor phase.

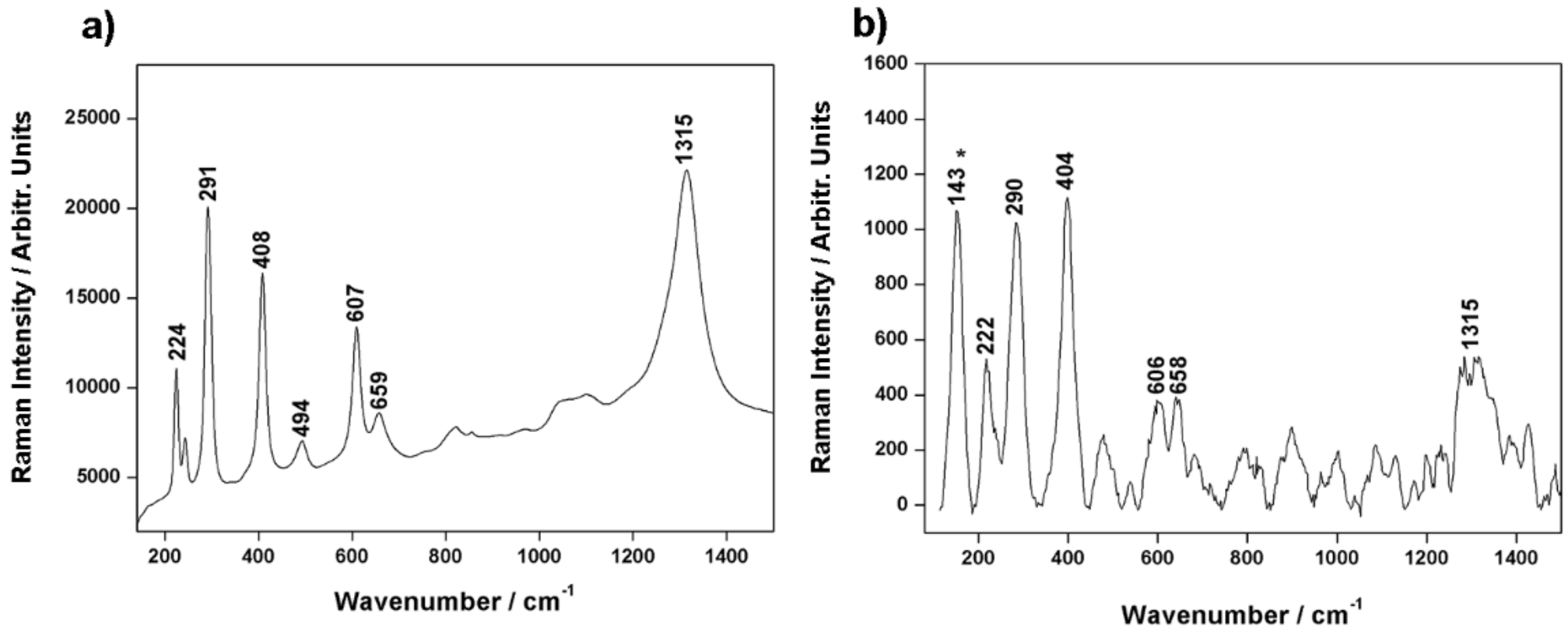

*Fig. 6: Hematite spectra collected from sample JD16-0005 by means of both microRaman system (a) and RLS ExoMars Simulator (b). * anatase contribution.*

*Table 5: Overview of Raman results collected from John Day Formation samples.*

| **Sample details** | | **Detected compounds** | | | | | | | | | | | | | |
|---|---|---|---|---|---|---|---|---|---|---|---|---|---|---|---|
| ID code | Sampling site | Feldspar | Pyroxene | Quartz | Hematite | Goethite | Olivine | Serpentine | Apatite | Anatase | Calcite | Carbon | Barium sulphate | Ilmenite | Clinoptilolite |
| JD16-0001 | Foree | | | | | | | | | ▥ | ▥ | | | | |
| JD16-0002 | Foree | ▤ | | ▩ | ▤ | | | | | ▩ | ▤ | | | | |
| JD16-0003 | Picture Gorge | ▤ | | | | | | | | | | ▩ | | | |
| JD16-0004 | Picture Gorge | ▩ | ▥ | | ▤ | | | | | ▥ | | | | | ▤ |
| JD16-0005 | Picture Gorge | ▩ | ▩ | | ▩ | | | | | ▥ | | | | | |
| JD16-0006 | Picture Gorge | ▩ | | | ▩ | | | | | | | | | | |
| JD16-0007 | Picture Gorge | ▩ | ▩ | | ▤ | | ▤ | | | ▥ | | | | | |
| JD16-0008 | Mascall Basin | ▤ | | | ▤ | | | | | ▥ | | | | | |
| JD16-0009 | Mascall Basin | ▤ | | | ▤ | | | | | ▩ | ▤ | | | | |
| JD16-0010 | Painted Hills | | | | ▩ | | | | | ▥ | | | | | |
| JD16-0011 | Painted Hills | ▩ | | | ▩ | | | | ▩ | | | ▥ | | ▤ | |
| JD16-0012 | Painted Hills | | | | ▩ | | | | | ▥ | | | | | |
| JD16.0013 | Painted Hills | | | | ▩ | | | | ▤ | ▥ | | | | | |
| JD16-0014 | Painted Hills | ▩ | | | ▤ | | | | | ▥ | | ▥ | | | |
| JD16-0015 | Painted Hills | ▤ | | ▥ | ▩ | | | | | ▥ | | ▥ | | | |
| JD16-0016 | Hancock Field Station | | | | ▩ | | | | | ▥ | | | | | |
| JD16-0017 | Hancock Field Station | ▤ | | ▥ | ▤ | | | | | ▥ | | | | | ▤ |
| JD16-0018 | Hancock Field Station | | | ▩ | | ▤ | | ▤ | | ▥ | | | ▤ | | |
| JD16-0019 | Hancock Field Station | | | ▩ | | ▤ | | ▤ | | ▥ | | | | | ▤ |
| JD16-0020 | Painted Hills | ▩ | | | ▩ | | | | ▩ | ▥ | ▩ | | | | |
| JD16-0021 | Painted Hills | ▩ | | | ▤ | | | | ▩ | ▥ | ▤ | | | | |
| JD16-0022 | Painted Hills | | | | ▤ | | | | | ▥ | | ▥ | | | |
| JD16-0023 | Painted Hills | ▤ | | | ▤ | | | | | ▥ | | ▥ | | | |
| JD16-0024 | Painted Hills | ▩ | | | ▤ | | | | | ▩ | | ▥ | | | |

Both microRaman and RLS simulator | RLS simulator | microRaman

### 3.2.4 Leka

As described elsewhere,[39] the chemical reactions produced by the iteration between ultramafic rocks and low temperature waters are considered to play a key role in the proliferation of chemolithotrophic microbes deep below the surface of Earth. In the past years, ultramafic rocks from the island of Leka (Norway) have been studied to determine the role played by these reactions in the proliferation of microbial communities on both Earth and Mars.[27,28] Thus, 17 samples were collected from several locations to extrapolate information about the weathering processes of ultramafic rocks. [40]

According to microRaman and RLS ExoMars Simulator results, olivine and pyroxene are the major mafic minerals composing Leka samples (excluding gabbroic and pillow lava analogues) pyroxene, while feldspars were only identified on samples LE16-0013 and LE16-0015.

As shown in Table 6, Raman analysis also provided molecular data demonstrating the occurrence of different alteration processes on most of the sampled materials. On one hand, the first 8 samples on Table 6 (from LE16-0001 to LE16-0008), together with the analogue LE16-0016, present serpentine (general formula $X_{2-3}Y_2O_5(OH)_4$, where X can be Mg, Mn, Fe, Ni, Al or Zn, while Y can be Si, Al or Fe. Main peaks at 230, 386, 690, 3685 and 3705 $cm^{-1}$ [41]) as the only alteration product. Serpentinization is a common metamorphic process of ultramafic rocks, involving the oxidation and hydrolyzation of olivine. In the case of sample LE16-0002 the serpentinization process was so intense to cause the complete transformation of olivine. Being a phyllosilicate, serpentine is considered a water-bearing mineral ($H_2O$ vibration features can be observed in the range from 3500 to 3800 $cm^{-1}$ of the two spectra represented in Fig. 7) of critical importance in the research of biomarkers in Mars.

On the other hand, additional degradation products were detected in the remaining samples. Calcite was clearly observed in samples LE16-0013, LE16-0014, LE16-0015 and LE16-0017, confirming the occurrence of carbonatation processes. Similarly, the detection of epidote ($Ca_2(Al,Fe)_2(SiO_4)_3(OH)$, main Raman peaks at 301, 337, 431, 453, 490, 572, 675, 872, 924, 1070 and 1093 $cm^{-1}$ [27]) in samples LE16-0009 and LE16-0014, as well as the detection of apatite in sample LE16-0013, proved the hydrothermal alteration of the sampled rocks.

As previously mentioned, LE16-0009, LE16-0011 and LE16-0014 samples were classified as gabbroic rocks. In spite of the fact that gabbroic samples are mainly composed of pyroxene and plagioclase, the latter was not detected on any of the collected materials. Furthermore it must be also underlined that, in the case of sample LE16-0009, the additional detection of a further phyllosilicate (chlorite, general formula $(Fe,Mg,Al)_6(Si,Al)_4O_{10}(OH)_8$, main peaks at 200, 549 and 671 $cm^{-1}$ [37]) confirmed the strong alteration of the selected material.

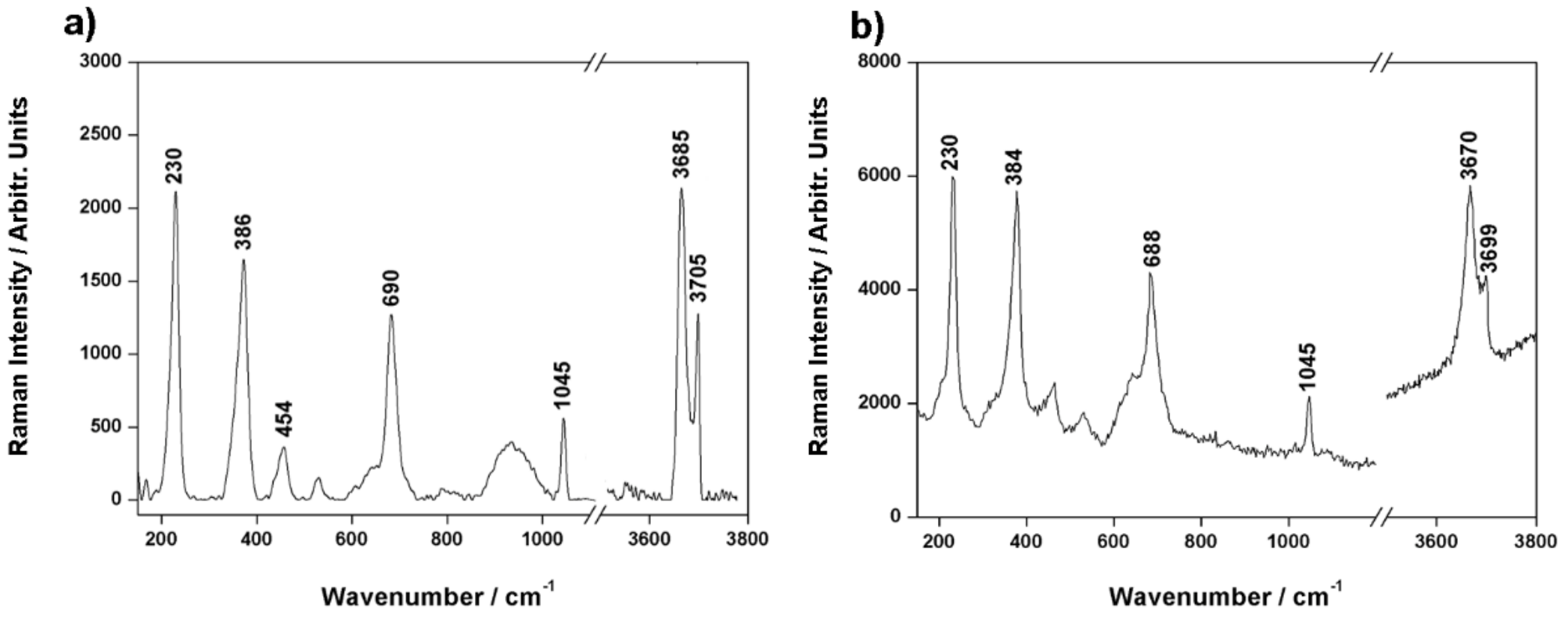

*Fig. 7: Serpentine spectra collected from sample LE16-0004 by means of both microRaman system (a) and RLS ExoMars Simulator (b).*

*Table 6: Overview of Raman results from Leka samples.*

| **Sample details** | | **Detected compounds** | | | | | | | | | | | | | |
|---|---|---|---|---|---|---|---|---|---|---|---|---|---|---|---|
| ID code | Sampling site | Feldspar | Pyroxene | Quartz | Magnetite | Hematite | Hornblende | Olivine | Serpentine | Apatite | Chlorite | Calcite | Epidote | Carbon | Ilmenite |
| LE16-0001 | Lauvhatten | | ▩ | | | | | ▩ | ▩ | | | | | | |
| LE16-0002 | Lauvhatten | | ▥ | | | ▤ | | | ▩ | | | | | | |
| LE16-0003 | Lauvhatten | | ▥ | | | ▤ | | ▩ | ▩ | | | | | | ▤ |
| LE16-0004 | At Skråen road | | ▥ | | | | | ▩ | ▩ | | | | | | |
| LE16-0005 | At Steinfjellet | | | | | | | ▩ | ▩ | | | | | | ▤ |
| LE16-0006 | At Steinfjellet | | | | | | | ▩ | ▩ | | | | | | |
| LE16-0007 | At Steinfjellet | | ▥ | | | | | ▩ | ▩ | | | | | | ▤ |
| LE16-0008 | At Pavillion | | ▩ | | | | | ▩ | ▩ | | | | | | |
| LE16-0009 | Aunkollen | | ▩ | | | ▤ | | | | | ▩ | | ▩ | | ▤ |
| LE16-0010 | Aunkollen | | | | | | | ▩ | ▩ | | | | | | |
| LE16-0011 | Aunkollen | | ▩ | | | | | | | | | | | | |
| LE16-0012 | Kvaløy | | | | | | | ▩ | ▩ | | | | | | |
| LE16-0013 | Madsøy | ▩ | | ▤ | | ▤ | ▩ | | | ▤ | | ▩ | | | |
| LE16-0014 | Solsem | | | | | | ▩ | | | | | ▩ | ▩ | ▤ | |
| LE16-0015 | Solsemhola | ▤ | ▩ | | | | | ▩ | ▩ | | | ▥ | | | ▤ |
| LE16-0016 | Moho | | ▥ | | ▤ | ▤ | | ▩ | ▩ | | | | | | |
| LE16-0017 | Moho | | | | ▤ | ▤ | | ▩ | ▩ | | | ▩ | | | |

▩ *Both microRaman and RLS simulator* ▥ *RLS simulator* ▤ *microRaman*

*3.2.5 Oslo rift*

The Tharsis region on Mars displays massive volcanoes and canyons formed during the Noachian period (4.1-3.7 billion years ago), which are believed to have formed due to extensional tectonics processes.[42] Accordingly, geological samples proceeding from the Oslo rift (Norway) were included in the PTAL database as terrestrial analogues of extensional tectonics rifting.[43] In detail, 3 weakly altered gabbro samples were collected from the region of Brattåsen (BR16-0001 and BR16-0002) and Ullernåsen (UL16-0001) and analyzed for comparison purposes.

Pyroxenes (Fig. 8), feldspars and calcite were identified as major compounds of Brattåsen samples. In addition to those quartz and apatite were detected on sample BR16-0001, while olivine and ilmenite were identified as minor compounds of sample BR16-0002.

With regards to sample UL16-0001 (Ullernåses), most Raman spectra displayed clear peaks around 475 and 510 cm$^{-1}$, suggesting that feldspars constitute the main mineralogical phase of the original rock. Of the 40 spectra obtained with the RLS ExoMars Simulator, half of them returned an intense peak at 1086 cm$^{-1}$, identifying calcite as the main alteration product of the sample. In addition to those, Raman data also confirmed the minor presence of additional compounds such as quartz, pyroxenes, olivine and apatite.

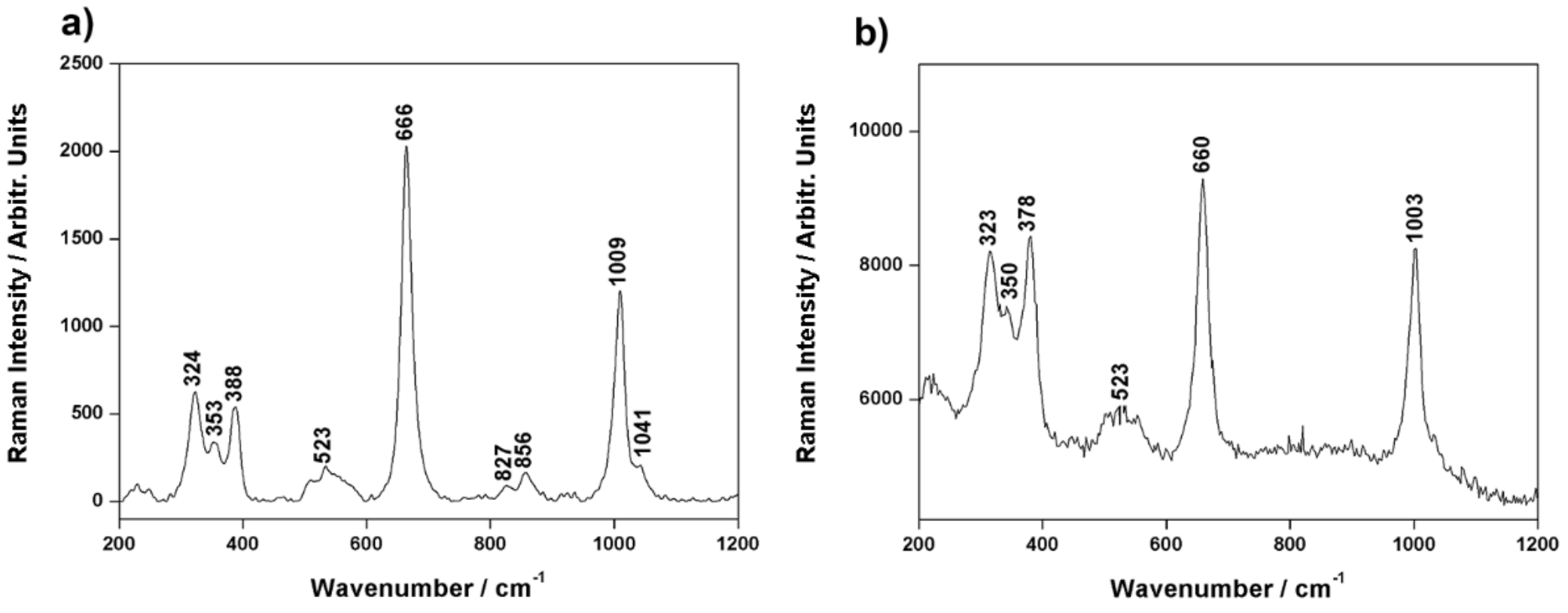

*Fig. 8: Pyroxene spectra collected from sample BR16-0001 by means of both microRaman system (a) and RLS ExoMars Simulator (b).*

*Table 7: Overview of Raman results from Oslo Rift samples.*

| Sample details | | Detected compounds | | | | | | | |
|---|---|---|---|---|---|---|---|---|---|
| ID code | Sampling site | Feldspar | Pyroxene | Quartz | Olivine | Apatite | Gypsum | Calcite | Ilmenite |
| BR16-001 | Brattåsen | ▩ | ▩ | ▤ | | ▩ | | ▤ | |
| BR16-002 | Brattåsen | ▤ | ▩ | | ▩ | | | ▩ | ▤ |
| UL16-001 | Ullernåsen | ▩ | ▩ | ▩ | | ▩ | ▤ | ▩ | |

▩ *Both microRaman and RLS simulator* ▥ *RLS simulator* ▤ *microRaman*

## 3.3 Impact craters

The Martian surface shows more than 43000 craters produced by the impact of extraterrestrial bodies with the planetary crust. The mineralogical study of the stratigraphic units composing craters walls and basins can provide important clues regarding the evolution of Martian geology and environment. In the light of the forthcoming ExoMars and Mars2020 missions, five terrestrial craters were considered for the PTAL project so as to evaluate the scientific capabilities of spectroscopic systems in detecting impact-related alteration products.

### *3.3.1 Chesapeake Bay*

The 85km wide Chesapeake Bay impact structure (CBIS) dates back to 35.3 million of years ago and is considered the best preserved wet-target crater on Earth.[44] The impact structure is buried between 200 and 500 meters beneath the surface and, according to previous stratigraphic studies, it was formed after a shallow marine impact on Eocene Virginian shelf.[45] In the light of the recent discovery of Martian craters having morphological features comparable to CBIS,[46] two samples were selected from the breccia of the Chesapeake Bay crater to evaluate the role that spectroscopic systems could play in the characterization of impact craters on Mars.

The two samples, named WH16-005 and WH16-014, were collected from a depth of 1397.2 and 1407.2 m respectively (ICDP-USGS Eyreville core), which correspond to the upper melt-rich section of the suevitic and lithic impact breccias.

Raman analysis of sample WH16-005 detected quartz, rutile, calcite, biotite and carbon as main mineralogical phases of the melted rock. However, no mineral amorphization induced by the impact was found. In the case of sample WH16-014, microRaman and RLS ExoMars Simulator analysis identified quartz as the main mineralogical phase of the analogue. By comparing the obtained quartz spectra with a reference standard, a clear shift of the main peak towards lower wavelengths (Fig. 9) was systematically observed. This displacement, which is induced by the amorphization of quartz, constitutes the analytical clue of the impact origin of the crater. Furthermore, in about 30% of quartz spectra two additional vibrational signals were detected at 230 and 416 $cm^{-1}$. These peaks corroborate the presence of cristobalite ($SiO_2$) a quartz polymorph that crystallizes at temperatures higher than 1450 ºC.

Further compounds were also identified including titanium- and iron-based oxides (rutile, ilmenite and hematite). Among minor phases, the detection of barite and siderite ($FeCO_3$, main peaks at 145, 277, 711, and 1086 $cm^{-1}$ [29]), minerals that can crystallize under hydrothermal conditions, provided clues about the occurrence of water-related alteration processes on the impact crater.

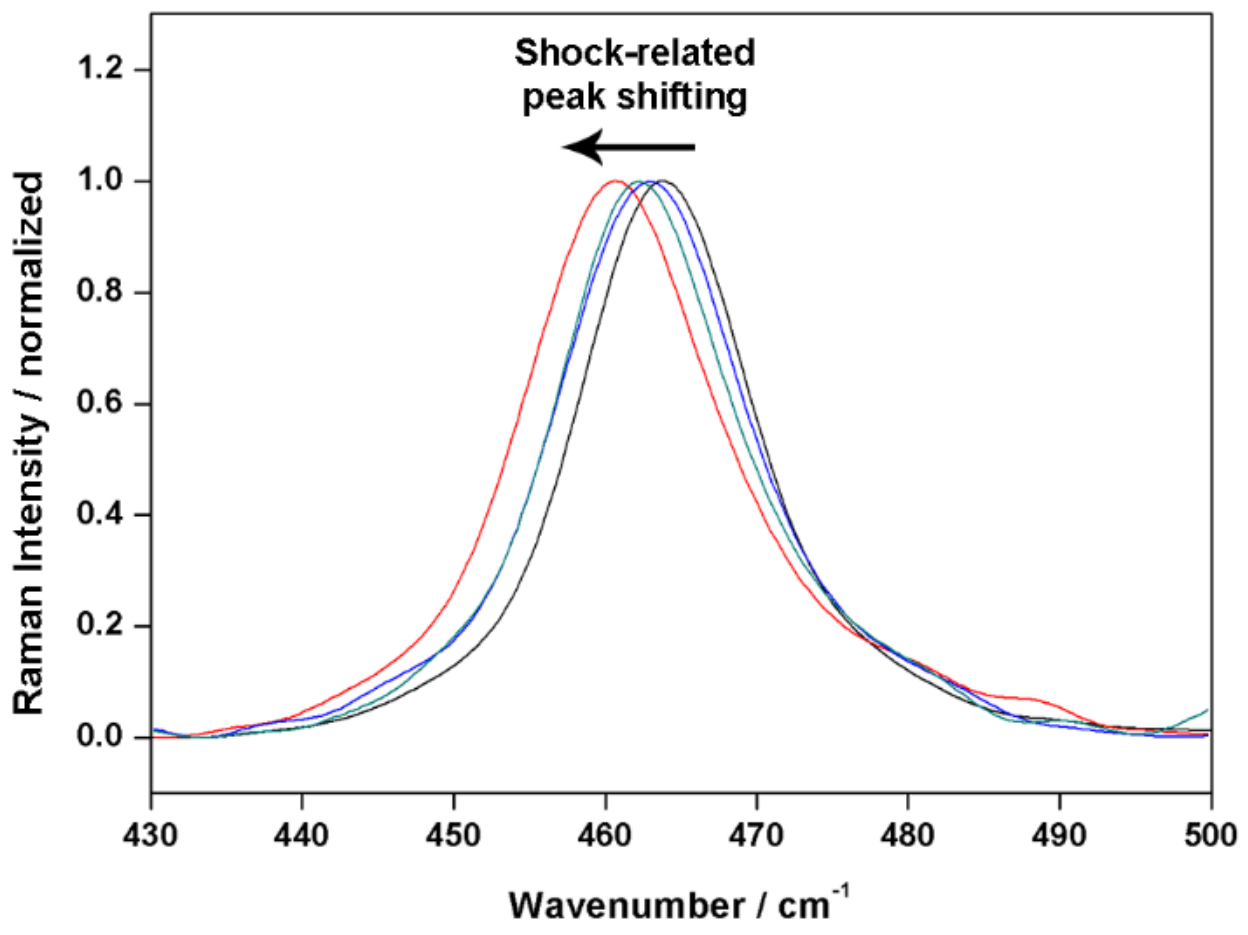

*Fig. 9: Comparison of Raman spectra collected by means of the microRaman system from different quartz grains of sample WH16-0014. The observed peak shifting is related to the shock-amorphization of the crystals caused by the impact.*

*3.3.2 Gardnos*

The 5 km wide crater of Gardnos (Norway) was produced between 900 and 400 million of years ago by the impact of a 250 m bolide.[47,48] For this project, 100 grams of melt-bearing impactite were sampled from central part of Gardnos structure and analyzed to identify the mineralogical amorphizations produced by the extreme pressures and temperatures released during impact. Most of the spectra obtained from sample GN16-001 displayed vibrational features that can be clearly assigned to feldspars. Concretely, the constant detection of two main vibrational peaks at 478 and 511, together with the clear identification of secondary signals at 287, 329, 409, 582, 763, 814 and 1123 $cm^{-1}$ (Fig. 10), proves the presence of K-rich feldspars (plagioclase). The vibrational features of amorphous carbon were also observed in most of the collected spectra. Beside plagioclase and carbon, a few crystals of quartz and anatase were also identified as minor mineralogical inclusions of the sample matrix. In the case of study, it must be underlined that the quartz signal (464 $cm^{-1}$) was often covered by one of the main peaks of plagioclase (478 $cm^{-1}$). For this reason, Raman analysis did not provide any definitive clue regarding the shock-induced amorphization of $SiO_2$.

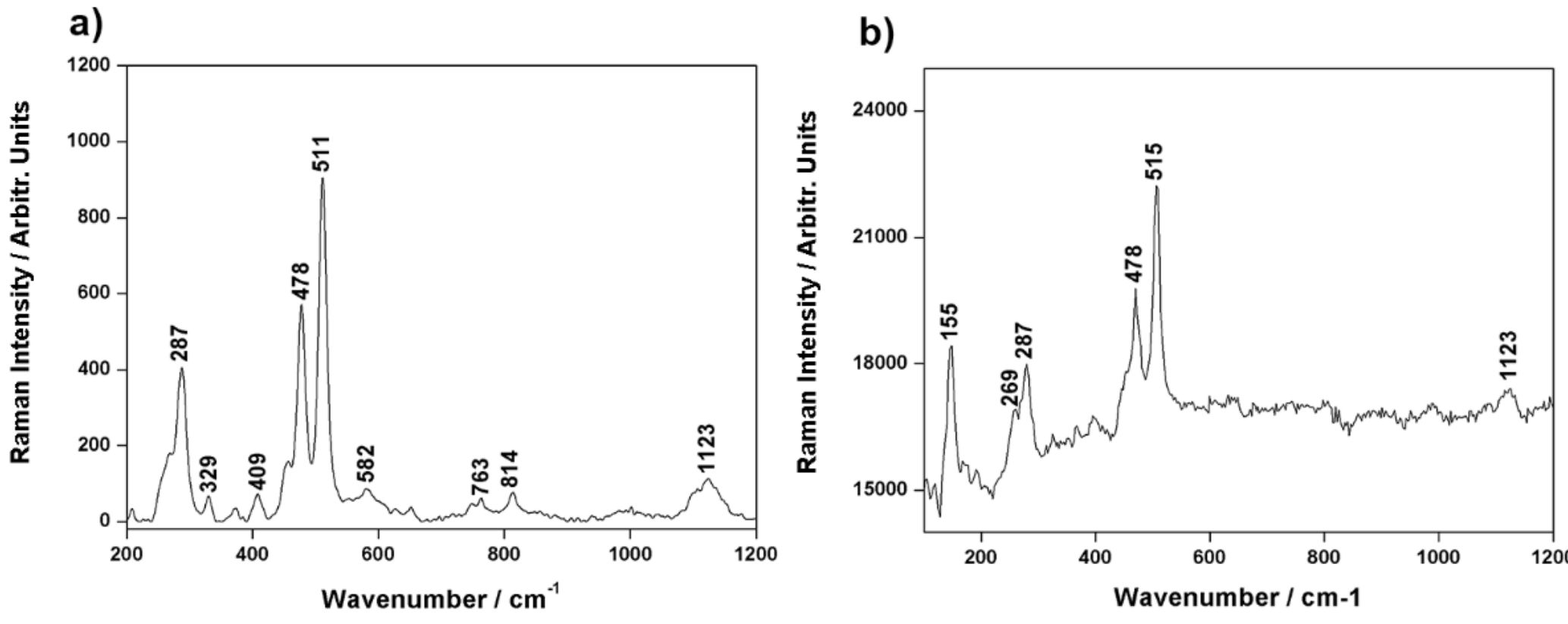

*Fig. 10: Plagioclase spectra collected from sample GN16-0001 by means of both microRaman system (a) and RLS ExoMars Simulator (b).*

*3.3.3 Vredefort*

The 300 km wide crater of Vredefort (South Africa) dates back to 2000 millions of years ago and it was produced by the impact of a 5 to 10 km wide bolide. As reported by L. Carporzen et al.,[49] the Vredefort crater represents the oldest impact structure discovered on Earth and is considered to be an optimal terrestrial analogue of large Martian craters. In the framework of the PTAL project, 130 grams of impact melt rocks were sampled from the Leeukop Quarry area to study the shock deformations suffered by target minerals as a consequence of the impact (above 30 GPa). In this case, the results obtained from the use of the RLS ExoMars Simulator perfectly fit with those provided by the microRaman system. As in the case of sample GN16-0001, most Raman spectra displayed the typical vibrational peaks of plagioclases. Beside K-feldspars, the characteristic signals of pyroxene were detected in 20% of the analysed spots. The main peak of quartz, detected in a few spectra, always appeared at 464 cm$^{-1}$. According to the work of P.F. Mc Millan et al.,[50] the collected material was therefore submitted to shock pressures below 12 GPa. Besides quartz, plagioclase and pyroxene, no additional minerals were detected, indicating a minimum or null post-impact alteration of the sampled material (Table 8).

*Table 8: Overview of Raman results from Impact Crater samples.*

| **Sample details** | | **Detected compounds** | | | | | | | | | | | | |
|---|---|---|---|---|---|---|---|---|---|---|---|---|---|---|
| ID code | Sampling site | Feldspar | Pyroxene | Quartz | Cristobalite | Hematite | Anatase | Rutile | Calcite | Siderite | Biotite | Carbon | Ilmenite | Barite |
| GN16-0001 | Gardnos | ▩ | | ▩ | | | ▥ | | | | | ▩ | | |
| VR16-0021 | Vredefort | ▩ | ▩ | ▩ | | | | | | | | | | |
| WH16-0005 | Chesapeake Bay | | | ▩ | | | | ▩ | ▩ | | ▤ | ▩ | | |
| WH16-0014 | Chesapeake Bay | ▩ | | ▩ | ▩ | ▩ | | ▩ | | ▥ | | | ▤ | ▩ |

▩ *Both microRaman and RLS simulator* ▥ *RLS simulator* ▤ *microRaman*

*3.3.4 Vista Alegre and Vargeão Dome*

As reported elsewhere.[51] 4 craters produced by the impact of a bolide against basaltic rocks were recently discovered in Brazil. Among them, the 12km wide impact structure of Vargeão Dome (124 million of years ago) was selected for the PTAL database as analogue of Martian basalt craters. The impact origin of this crater was confirmed in 1993, when Hachiro et al.,[52] identified evidence of planar deformation features (PDF) in quartz and feldspar crystals collected from the crater basin. In this case, two analogues (VO16-0001 and VO16-0002) were sampled from polymictic breccia rocks. According to Raman data, feldspars constitute the main mineral phase of both samples. Further original minerals such as anatase and quartz were also detected in the two analogues, while the presence of pyroxenes was only observed on sample VO16-0001. In addition to those, a few alteration products were also identified. Hematite was observed in both samples, while apatite and calcite were only punctually identified on samples VO16-0001 and VO16-0002 respectively.

In close proximity to Vargeão Dome, the Vista Alegre impact structure was also considered for the PTAL database as basaltic-target crater. This 9.5 km wide structure was discovered in 2004 and it is estimated to have been formed about 115 million of years ago.[53] As in the previous case of study, feldspars, pyroxenes and quartz were clearly characterized as the main original compounds of the polymictic breccia sample (VA16-0001). The only discrepancy consists in the presence of ilmenite (instead of rutile) as main Ti-based mineral. The further detection of hematite, calcite and apatite (Fig. 11) proves a moderate rock alteration.

Sample VO16-0003, sampled nearby the Vargeão Dome crater, was included in the PTAL database to provide future users with a reference for unshocked basalt from the Serra Geral Formation (Brazil). By comparing Raman data with the results summarized above, the detection of higher amounts of pyroxene crystals as well as the lack of alteration compounds such as hematite and calcite must be underlined.

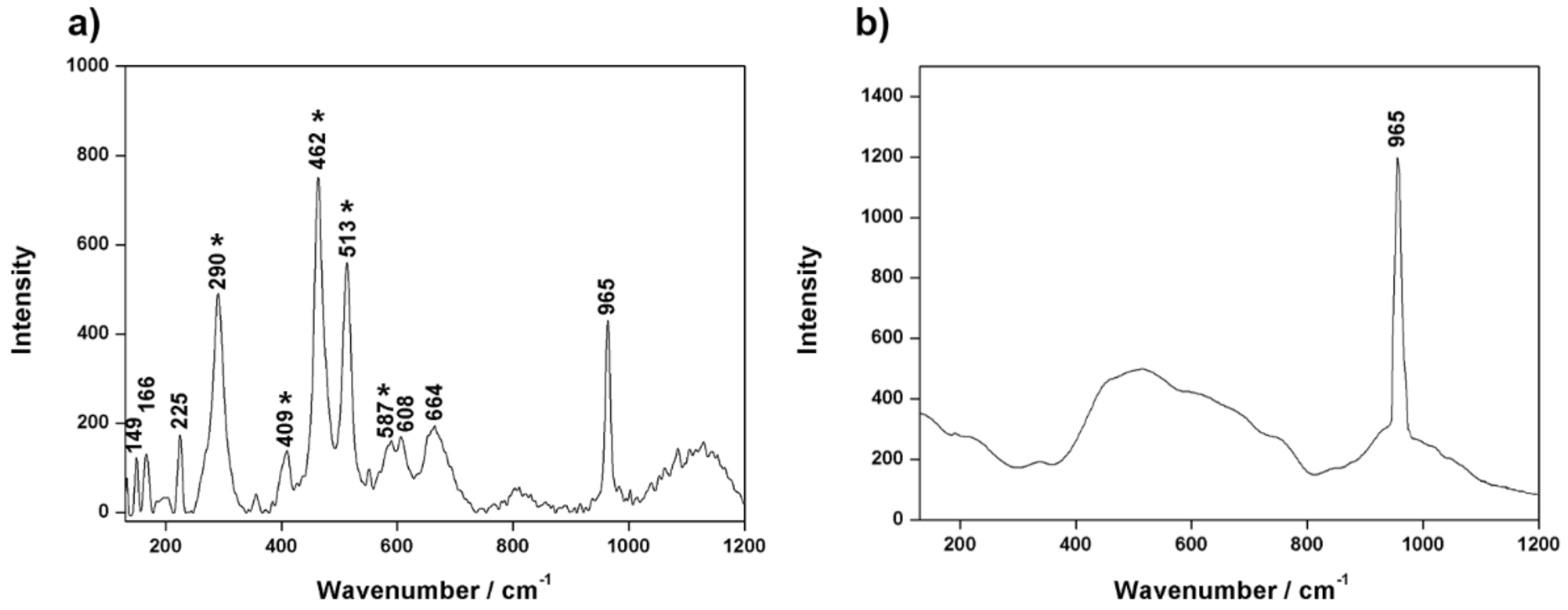

*Fig. 11: Apatite spectra collected from sample VO16-001 by means of both microRaman system (a) and RLS ExoMars Simulator (b). *feldspar contribution.*

*Table 9: Overview of Raman results from Vista Alegre and Vargeão Dome samples.*

| Sample details | | Detected compounds | | | | | | | | | |
|---|---|---|---|---|---|---|---|---|---|---|---|
| ID code | Sampling site | Feldspar | Pyroxene | Quartz | Hematite | Magnetite | Anatase | Apatite | Calcite | Gypaum | Ilmenite |
| VO16-001 | Vargeao Dome | ▒ | ▥ | ▤ | ▤ | | ▥ | ▒ | | ▥ | |
| VO16-002 | Vargeao Dome | ▤ | | ▥ | ▤ | | ▥ | | ▒ | | |
| VA16-001 | Vista Allegre | ▒ | ▤ | ▥ | ▤ | ▤ | | ▤ | ▥ | | |
| VO16-003 | Vargeao Dome | ▒ | ▒ | ▤ | | | | ▤ | | | ▤ |

▒ *Both microRaman and RLS simulator* ▥ *RLS simulator* ▤ *microRaman*

**Conclusion**

This work presents the whole set of Raman data (over 4500 spectra) that have been collected to feed the PTAL database. In a general perspective, it was confirmed that Raman spectroscopy is able to reveal the complex mineralogical composition of a wide variety of terrestrial analogues, providing results in accordance with analytical studies presented in previous works.

Besides the multispectral database, the PTAL project will also provide users with spectra-treatment tools and physical access to the 94 terrestrial analogues. In this sense the PTAL information system aims to become a cornerstone tool for the scientific community interested on deepening the knowledge of geological processes occurred on Mars and other extraterrestrial bodies.

Beyond the PTAL project, the data summarized in this works also helps to shed light on the potential contribution of Raman spectroscopy to the mineralogical characterization of extraterrestrial bodies. On one hand, the characteristic vibrational Raman spectra provided by minerals (and potential biomarkers) are characterized by the presence of very sharp and well separated peaks, which facilitate the proper characterization of the analysed material even in the presence of complex mixtures. On the other hand, being the spot of analysis of micrometric size, a multipoint procedure enables the detection of minor and trace compounds that cannot be easily detected by other techniques.

In the light of the forthcoming ExoMars mission, the comparison figures and tables presented in this work proves that the spectra obtained by means of the RLS ExoMars Simulator are qualitatively comparable with those obtained with the microRaman laboratory system, which is an excellent indicator of the scientific capabilities of the RLS spectrometer that will land on Mars in 2020.

**Acknowledgments:**

This project is financed through the European Research Council in the H2020- COMPET-2015 programme (grant 687302) and the Ministry of Economy and Competitiveness (MINECO, grants ESP2014-56138-C3-2-R and ESP2017-87690-C3-1-R).